\documentclass[
  ,final            
  ]
  {aipproc}
\layoutstyle{6x9}

\usepackage[english]{babel}
\usepackage{amsmath,amssymb,amsfonts,amstext}
\usepackage{slashed}

\begin{document}

\title{Regge-plus-resonance predictions for neutral-kaon photoproduction from the deuteron}

\author{P.~Vancraeyveld\footnote{Email address: Pieter.Vancraeyveld@UGent.be}}{
	address={Department of Physics and Astronomy, Ghent University, Proeftuinstraat 86, B-9000 Gent, Belgium}}
\author{L.~De~Cruz}{
	address={Department of Physics and Astronomy, Ghent University, Proeftuinstraat 86, B-9000 Gent, Belgium}}
\author{J.~Ryckebusch}{
	address={Department of Physics and Astronomy, Ghent University, Proeftuinstraat 86, B-9000 Gent, Belgium}}

\begin{abstract}
We present a Regge-inspired effective-Lagrangian framework for neutral-kaon photoproduction from the deuteron.
Quasi-free kaon production is investigated using the Regge-plus-resonance elementary operator 
within the relativistic plane-wave impulse approximation.
The Regge-plus-resonance model was developed to describe photoinduced and electroinduced charged-kaon production off protons.
We show how this elementary operator can be transformed to account for the production of neutral kaons from both protons and neutrons.
Our results compare favourably to the sole ${^2\text{H}}(\gamma,K^0)YN$ dataset published to date.
\end{abstract}


\maketitle


Electromagnetic production of strangeness plays a prominent role in current efforts to chart the excitation spectrum of the nucleon.
Since the production mechanism inevitably involves quark-antiquark components of the nucleon's sea,
it offers the opportunity to improve on our understanding of the strong interaction in the confinement regime.
Six different kaon photoproduction channels exist which can be treated within a single theoretical framework.
The current kaon photoproduction database is heavily dominated by the $p(\gamma,K^+)\Lambda$ and $p(\gamma,K^+)\Sigma^0$ channels.
A comprehensive survey of the four remaining reactions, 
namely $n(\gamma,K^0)\Lambda$, $p(\gamma,K^0)\Sigma^+$, $n(\gamma,K^0)\Sigma^0$ and $n(\gamma,K^+)\Sigma^-$,
would provide useful complementary information to elucidate the strangeness-production reaction mechanism.
In this respect a trustworthy description of kaon production from the deuteron is of chief importance,
for the deuteron's weak binding energy makes it a prime neutron target.
In this contribution, we focus on the semi-inclusive production of neutral kaons from a deuteron target 
as it is the only channel for which data is available~\cite{LNSiso4}.

Whilst strangeness production is an important tool in nucleon spectroscopy
and is in part mediated by nucleon-resonance exchange,
the smooth energy dependence of the measured observables hints at a dominant role for the background, i.e.\ non-resonant, processes.
The Regge-plus-resonance (RPR) approach to kaon production seeks to decouple the determination of the coupling constants for the background and the resonant diagrams.
This has resulted in a hybrid model which accounts for electromagnetic kaon production 
from threshold up to $E_{\gamma}=16\;\text{GeV}$~\cite{RPRlambda,RPRsigma,RPRelectro,RPRneutron}.
The non-resonant contributions to the amplitude are efficiently modelled 
in terms of $K^+(494)$ and $K^{\ast+}(892)$ Regge-trajectory exchange in the $t$-channel~\cite{GuidalPhotoProdKandPi}.
The three coupling constants can be unambiguously determined given sufficient high-energy ($E_{\gamma}\gtrsim4\;\text{GeV}$) data~\cite{RPRbayes}.
The Regge amplitudes are supplemented with a selection of $s$-channel resonance-exchange diagrams,
whose parameters are optimised to data in the resonance-region ($E_{\gamma}\lesssim4\;\text{GeV}$) while keeping the background amplitude anchored.
For the $p(\gamma,K^+)\Lambda$ reaction, a set of established nucleon resonances turned out to be insufficient.
The addition of a $N(1900)D_{13}$ resonance made it possible to accurately describe both photo- and electroproduction data~\cite{RPRlambda,RPRelectro}.
The $p(\gamma^{(\ast)},K^+)\Sigma^0$ data, on the other hand,
is properly described within the RPR framework considering only established nucleon and delta resonances~\cite{RPRsigma,RPRelectro}.
Accordingly, there is no direct hint for missing resonances in that channel.

In the $K^+\Lambda$ and $K^+\Sigma^0$ production channels, sufficient data is available to constrain the free parameters of the production operator.
For the remaining channels ($K^0\Lambda$, $K^0\Sigma^+$, $K^0\Sigma^0$, $K^+\Sigma^-$) the small amount of data does not allow one to tune the model parameters.
In the strong interaction vertex of the considered Feynman diagrams, 
the coupling constants can be transformed assuming SU(2) isospin symmetry.
Converting the electromagnetic coupling constants requires experimental input.
For the production of neutral kaons two alterations to the background amplitude are required.
First, the $K(494)$-exchange $t$-channel diagram does no longer contribute
and second, a value for the magnetic transition moment of the $K^{\ast0}(892)$ to the $K^0(494)$ is needed.
Optimizing the RPR model to the $p(\gamma,K^0)\Sigma^+$ dataset~\cite{SAPHIRsigma0,Castelijns,PHDcarnahan}, 
we obtain $\kappa_{K^{\ast0}(892)K^{0}(494)}/\kappa_{K^{\ast+}(892)K^{+}(494)}=0.05\pm0.01$.
For kaon photoproduction on a neutron, 
the nucleon-exchange diagrams need to be transformed using photocoupling helicity amplitudes,
which are extracted from pion production reactions, amongst others.
Confronting RPR predictions with $n(\gamma,K^+)\Sigma^-$ data has demonstrated that this approach gives fair results,
albeit with considerable uncertainties that stem from the experimental errors on the employed helicity amplitudes~\cite{RPRneutron}.
For the Reggeized background, its three parameters can be transformed adopting only isospin symmetry arguments, yielding robust predictions.

In order to embed the RPR elementary operator in the nuclear medium, we invoke the impulse approximation.
For now, we neglect final-state interactions and assume plane waves for the final-state particles.
The fully Lorentz-invariant expression for the reaction amplitude is
\begin{multline}\label{eq:matrixElement}
 \mathcal{M}^{\lambda_{\gamma},\lambda_D}_{\lambda_Y,\lambda_N}(\gamma D\rightarrow KYN) =\\
 \frac{1}{\sqrt{2}}
  \overline{u}(\vec{p}_Y,\lambda_Y)
   \epsilon_{\lambda_{\gamma}}\cdot\hat{J}_{\text{elem}}\left(\gamma N'\rightarrow KY\right)
   \frac{m_N+\slashed{p}_D-\slashed{p}_N}{m_N^2-\left(p_D-p_N\right)^2}
   \epsilon_{\lambda_D}\cdot\Gamma_{\text{\tiny Dnp}}(p_N,p_D)
   \mathcal{C} \overline{u}^T(\vec{p}_N,\lambda_N)\,,
\end{multline}
where the factor $\sqrt{2}$ stems from isospin factors and the fact that the production operator acts on a single nucleon. 
The $\lambda_i$ are the spin projections of the initial- and final-state particles 
and $\mathcal{C}$ is the charge conjugation matrix. 
Since the spectator nucleon (with momentum $p_N$) is on-mass shell, the covariant $Dnp$-vertex $\Gamma_{\text{\tiny Dnp}}$ is defined by~\cite{BlankenbeclerCook}
\begin{equation}
 \Gamma_{\text{\tiny Dnp}}^{\mu}(p_N,p_D) = 
  F(|\vec{p}|)\gamma^{\mu} - \frac{G(|\vec{p}|)}{m_N}p^{\mu}
  -\frac{m_N-(\slashed{p}_D-\slashed{p}_N)}{m_N}
    \left( H(|\vec{p}|)\gamma^{\mu} - \frac{I(|\vec{p}|)}{m_N}p^{\mu} \right)\,,
\end{equation}
with $p = \frac{1}{2}p_D-p_N$.
The four scalar form factors $F$, $G$, $H$ and $I$ can be expressed in terms of the $s$-, $p$- and $d$-wave components of the deuteron wave function~\cite{BuckGross}.
In our calculations we will adopt the relativistic wave functions obtained with the WJC-1 solution presented in Refs.~\cite{GrossWJC,GrossWJCparam}.

\begin{figure}
 \centering
 \includegraphics[width=.6\textwidth]{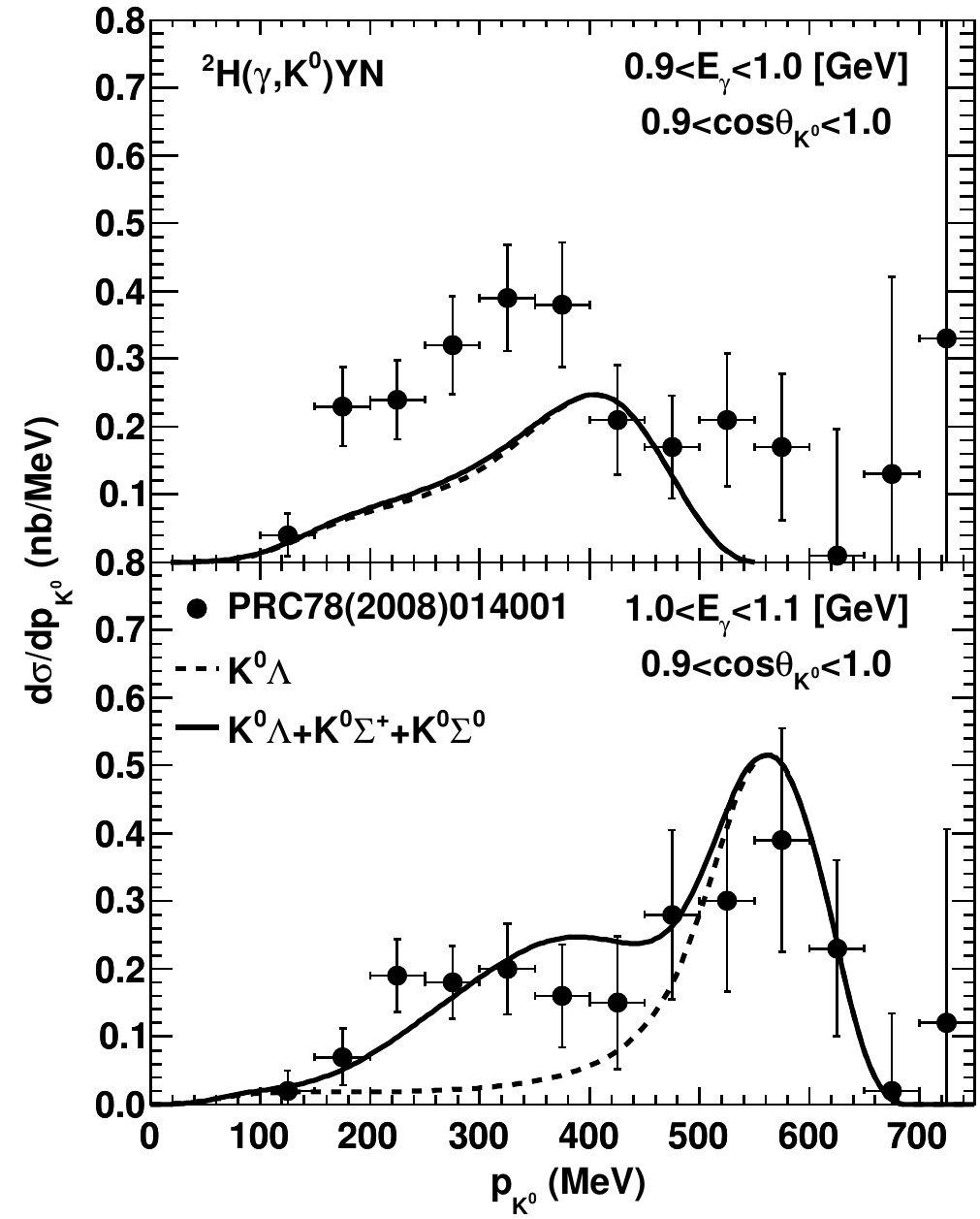}
  \caption{
The semi-inclusive ${^2H}(\gamma,K^0)YN$ differential cross section
as a function of the kaon momentum $p_K$
after integrating over $0.9\leq\cos\theta_{K^0}\leq1$.
The model results in the upper (lower) panel have been calculated at $E_{\gamma}=950 (1050)\,\text{MeV}$.
The solid line shows the full result, 
whereas the dashed line singles out the contribution of the ${^2H}(\gamma,K^0)\Lambda p$ reaction channel.
The data are from Ref.~\cite{LNSiso4}.}
 \label{fig:vsdata}
\end{figure}

Within the relativistic plane-wave impulse approximation~(RPWIA), 
the photoproduction of a neutral kaon from a deuteron target can proceed in three different elementary ways.
For two of these, $n(\gamma,K^0)\Lambda$ and $n(\gamma,K^0)\Sigma^0$, the incoming photon interacts with a bound neutron,
and no data is available to tune the RPR model.
In figure~\ref{fig:vsdata}, we compare the RPR predictions with the data obtained using the NKS at the Laboratory of Nuclear Science at Tohoku University~\cite{LNSiso4}.
They have measured the semi-inclusive neutral-kaon photoproduction cross section from deuterium in two photon-energy bins close to threshold.
The results are presented as a function of the momentum of the outgoing kaon $p_K$
and for a range of scattering angles $\theta_{K^0}$ in the forward direction:
\begin{multline}\label{eq:incXsection}
 \frac{d\sigma}{dp_{K^0}} =\\
 \sum_{Y=\Lambda,\Sigma^0,\Sigma^+}
 \int_{0.9}^1 \cos\theta_{K^0}\int d\Omega_Y^{\ast}
  \frac{1}{32(2\pi)^4}\frac{|\vec{p}_Y^*||\vec{p}_{K^0}|^2}{M_D E_{\gamma} E_{K^0} W_{YN}}
  \frac{1}{6} \sum
           \left| \mathcal{M}^{\lambda_{\gamma},\lambda_D}_{\lambda_Y,\lambda_N}(\gamma D\rightarrow K^0YN) \right|^2\,.
\end{multline}
In this expression all quantities are expressed in the laboratory frame, except for those marked with an asterisk,
and $W_{YN}$ is the invariant mass of the final hyperon-nucleon system.
In the lowest energy bin, where the $\Sigma$ production channels have not opened yet, the RPR model underpredicts the data by a factor of two approximately.
At the second photon energy, on the other hand, the description is far more satisfactory,
since both the $\Lambda$ and $\Sigma$ quasi-elastic peaks are well reproduced.

In summary, we have presented results for semi-inclusive neutral-kaon photoproduction from the deuteron within the RPWIA.
For the elementary process, we employ the Regge-plus-resonance model, which has been optimised against recent photoproduction data.
We adopt symmetry considerations to transform this elementary operator to account for the production of neutral kaons from both protons and neutrons.
With this technique, we obtain predictions for ${^2\text{H}}(\gamma,K^0)YN$ that describe the data fairly well.
Soon, things will improve on the experimental side 
with experiments at Jefferson Lab presently being analysed and others scheduled to run in the near future.
At Tohoku University, new data has been taken using an improved spectrometer with a larger acceptance~\cite{KandaPrivate}.
We intend to extend our formalism and investigate the role of hyperon-nucleon final-state interactions on exclusive and inclusive observables.
Besides making our model more complete,
this will give us access to the elusive hyperon-nucleon interaction
by focussing on kinematic regions where one expects important contributions from hyperons rescattering off the spectator nucleon.


\begin{theacknowledgments}
This work was supported by the Research Foundation -- Flanders (FWO) and the research council of Ghent University.
\end{theacknowledgments}

\bibliographystyle{aipproc}
\bibliography{../../bibliography/general}

\end{document}